# In Silico Trials for Sex-Specific patient Inclusion Criteria in Cardiac Resynchronization Therapy: Advancing Precision in Heart Failure Treatment

Shuang Qian[‡], Devran Ugurlu[§], Elliot Fairweather[*], Richard E Jones[‡, #, ^], Hassan Zaidi[*], Sanjay Prasad[‡,#], Brian P Halliday[‡, #], Daniel J Hammersley[*,#,&], Gernot Plank[†,1], Edward Vigmond[‖, ʃ], Christopher A Rinaldi[*], Alistair Young[*], Pablo Lamata[*], Martin Bishop[*], Steven Niederer[§,‡]

[‡]National Heart and Lung Institute, Imperial College London, London, UK
[§]Alan Turing Institute, British Library, 96 Euston Rd, London, UK
[*]Department of Biomedical Engineering, School of Imaging Sciences and Biomedical Engineering, Kings College London, London, UK
[#]Cardiovascular Magnetic Resonance Unit, Royal Brompton and Harefield Hospitals, Guy's and St Thomas' National Health Service Foundation Trust, London, UK
[^]Anglia Ruskin University, Chelmsford, UK
[†] Gottfried Schatz Research Center, Division of Biophysics, Medical University of Graz, Graz, Austria
[1]BioTechMed-Graz, Graz, Austria
[‖] University of Bordeaux, CNRS, Bordeaux, Talence, France
[ʃ] IHU Liryc, Bordeaux, Talence, France
[^]Anglia Ruskin University, Chelmsford, UK
[&]King's College Hospital NHS Foundation Trust

Corresponding address: Dr S Qian, 3rd Floor, ICTEM building, Hammersmith Campus, London, W12 0NN, UK (s.qian23@imperial.ac.uk)

# Abstract

Cardiac resynchronization therapy (CRT) guidelines are based on clinical trials with limited female representation and inconsistent left bundle branch block (LBBB) definitions. Conventional QRS duration (QRSd) criteria show variable diagnostic accuracy between sexes, partly due to differences in heart size and remodeling. We evaluated the influence of sex, heart size, LBBB, and conduction delay on QRSd and assessed the diagnostic performance of conventional and indexed QRSd criteria using a population-based modelling approach. Simulated QRSd were derived from electrophysiological simulations conducted in 2627 UK Biobank healthy participants and 359 patients with ischemic heart disease, by modelling LBBB and normal activation combined with/without conduction delay. QRSd criteria under-selected LBBB females and over-selected non-LBBB patients. Indexing by LVEDV and LV mass reduced sex disparities but increased the over-selection in non-LBBB patients. Height-indexed QRSd effectively resolved sex differences and maintained low non-LBBB selection rates, demonstrating superior performance and potential for more equitable CRT selection.

# Introduction

Cardiac Resynchronization Therapy (CRT) is a treatment that improves the cardiac function in patients with heart failure and left bundle branch block (LBBB). A key clinical biomarker used for selecting eligible patients is QRS duration—a reading on electrocardiograms (ECGs) reflecting the time taken for ventricular activation, important for assessing the health of the heart's conduction system. While CRT has been widely used and helped many patients, about 30% of patients do not show expected clinical improvements[1]. One factor that may contribute to this variation in outcomes is biological sex. While some studies, such as CARE-HF (Cardiac Resynchronization in Heart Failure)[2], reported no significant sex differences, most clinical evidence suggests that women demonstrate higher response rates to CRT, compared to men, particularly at shorter QRS durations[3].

The mechanisms underpinning these sex differences in CRT response remain poorly understood. Previous clinical studies have proposed that differences in heart sizes and QRS duration may partly explain the observed variation[4]. This was supported by our prior large-scale modelling study[5] which showed how the shorter QRS duration in women can be entirely explained by smaller heart sizes, without intrinsic differences in myocardial conduction velocity. Various indexed QRS duration criteria, such as by cardiac structural metrics including heart volumes and mass, or body habitus measures including height, have been proposed to resolve these differences and improve patient selection. But these studies have largely been examined in small clinical cohorts and only considered patients who met the conventional/unindexed QRS duration (QRSd) inclusion criteria[3]. As a result, such analyses may be biased toward showing that women who meet both conventional and indexed criteria benefit from CRT, while overlooking potential responders—both men and women—who meet indexed criteria but not conventional criteria.

In addition to structural remodelling, functional remodelling also plays a vital role in varying QRS duration in CRT patients/candidates. Functional remodelling can be generally categorised by two pathological processes: (1) left bundle branch block (LBBB), resulting in electrical desynchrony, and(2) slowed myocardial conduction due to the presence of scarring or myocardial fibrosis, ionic remodeling or gap junction down-regulation[4], which can both lead to the conduction delay and manifest as QRS duration prolongation. Previous studies demonstrated that patients with LBBB generally derive greater benefits from CRT compared to those without it[6]. To identify LBBB patients, various criteria based on QRS morphology in the electrocardiograms have been proposed, which have led to considerable variability in patient stratification[7], while the QRS duration criteria remain the gold-standard clinical biomarker for identifying CRT patients. However, the relative impacts of these distinct pathological substrates, or their combinations on QRS duration in different sexes, remain poorly understood. Furthermore, it is also unknown how these distinct pathological substrates affect the diagnostic efficacy of the current conventional QRS duration and indexed QRS duration in different sexes. Here we aim to predict whether LBBB patients can be effectively stratified using only QRS duration-based criteria, rather than relying on the complex and inconsistent ECG morphology-based criteria.

Computational models offer a unique platform for dissecting the impact of different pathological changes on QRS duration in a controlled manner. By leveraging patient-specific anatomical models derived from clinical imaging, it is possible to prospectively evaluate the diagnostic performance of various conventional or indexed QRS duration criteria under different pathological scenarios. A key limitation of prior studies is their focus on patients who already met conventional QRS duration criteria, rather than considering a broader population. Here, we introduce a novel approach using virtual cohorts, which allows for the exploration of a wide distribution of cases, including those who may not have met conventional inclusion criteria.

In this study, we hypothesize that the sex difference in CRT response is influenced by sex-dependent diagnostic accuracy for patients with different pathological changes, combined with the known factor of heart size differences. To test this, we prospectively evaluate the efficacy of conventional QRSd and indexed QRSd criteria for stratifying virtual LBBB and non-LBBB patients in different sexes under varying degrees of pathological changes. The simulated scenarios include LBBB and normal activation, in combination with/without slow myocardial conduction. These simulation scenarios were applied to patient-specific models developed from two large-scale cohorts: (1) healthy volunteers from the UK biobank (UKBB), representing anatomically healthy hearts, and (2) patients with ischemic heart disease (IHD), representing diseased anatomies with structural modelling. By expanding beyond conventional inclusion criteria and leveraging population-level models, this study provides new insights into the sex-specific efficacy of CRT selection criteria and their implications for optimizing patient outcomes.

# Results

## Basic patient characteristics

Table 1 summarizes participant characteristics for the healthy cohort (free from cardiovascular disease (CVD)) from UKBB and the IHD cohort. Females comprised 57% of the healthy cohort and 12.5% of the IHD cohort. Significant sex differences were observed in all cardiac structural metrics such as left ventricular end-diastolic volume (LVEDV) ($P < 8.1 \times 10^{-6}$), with the mean LVEDV and LV mass in IHD cohort being approximately 30% larger than those in the healthy cohort. Left ventricular ejection fraction (LVEF) differences by sex were non-significant in both cohorts, although the mean LVEF was lower in the IHD cohort (male: 45.2±16.0% vs 49.7±16.0%) compared to the healthy cohort (male: 55.0±6.1% vs 57.6±5.4%).

QRS durations measured at sinus rhythm were similar for both sexes in the healthy cohort (male: 93.2±23.8 ms vs 95.3±30.8 ms, P=0.7). However, in the IHD cohort, females exhibited lower QRS durations than males (100.8±26.5 ms vs 106.9.2±22.5 ms, P=0.01).

|  | UKBB (n=2627) | | | IHD (n=359) | | |
| --- | --- | --- | --- | --- | --- | --- |
|  | Male(n=1125) | Female (n=1502) | P value | Male(n=314) | Female (n=45) | P value |
| Age (year) | 61.2 ± 7.6 | 60.3 ± 7.3 | 0.001 | 64.2 ± 10.0 | 63.0 ± 9.1 | 0.34 |
| Weight (kg) | 83.0 ± 13.0 | 68.4 ± 12.3 | $9.8 \times 10^{-163}$ | 84.2 ± 15.4 | 70.4 ± 12.0 | $1.8 \times 10^{-8}$ |
| Height (cm) | 176.5± 6.4 | 163.1± 6.3 | $< 2 \times 10^{-308}$ | 174.2± 8.3 | 162.5± 6.8 | $3.9 \times 10^{-17}$ |
| BMI (kg/m^2) | 26.6±3.7 | 25.7±4.4 | $1.9 \times 10^{-14}$ | 27.8±4.9 | 26.7±4.6 | 0.21 |
| Left ventricular end-diastolic volume (LVEDV/mL) | 156.1±29.8 | 119.1±19.9 | $9.5 \times 10^{-218}$ | 222.3±81.5 | 171.2±66.4 | $8.1 \times 10^{-6}$ |
| Right ventricular end-diastolic | 158.7±29.4 | 114.8±20.3 | $4.3 \times 10^{-265}$ | 157.3±42.3 | 120.6±38.5 | $2.3 \times 10^{-8}$ |
| volume (RVEDV/ mL) | | | | | | |
| Left ventricular mass (LV mass/g) | 127.3±20.2 | 91.1±13.1 | $< 2 \times 10^{-308}$ | 162.2±51.3 | 123.8±42.7 | $3.6 \times 10^{-7}$ |
| Left ventricular ejection fraction (LVEF/ %) | 55.0±6.1 | 57.6±5.4 | 0.46 | 45.2±16.0 | 49.7±16.0 | 0.11 |
| QRS duration measured at sinus rhythm | 92.5±12.5 | 84±11.3 | $8 \times 10^{-90}$ | 106.9.2±22.5 | 100.8±26.5 | 0.01 |

Table 1: Basic characteristics of the healthy cohort from the UK biobank (n=2627) and the clinical cohort of patients with IHD (n=359).

## Simulated QRS durations in different pathological scenarios

Four scenarios were simulated for each patient-specific anatomical model and the resultant QRSds were calculated. Figure 1A showed QRSds simulated using healthy anatomies from UKBB with normal and LBBB activation. A 25% increase in QRSd was observed with LBBB activation (78.04±5.92 vs 97.56±7.55 ms, P<0.0001). Males exhibited longer QRSds than females in both groups (normal: 82.1±4.9 vs 75±4.7 ms and LBBB: 102.2±7.2 vs 93.7±5.1 ms, both P<0.0001). In scenarios involving additional slowed myocardial conduction (Figure 1B), QRSds were further prolonged (120.1±9.6 vs 145.5±12.9 ms, P<0.0001), with preserved sex-specific differences (healthy: 126.6±8.6 (male) vs 115.2±7.1 ms and LBBB: 153.4±12.9 (male) vs 139.5±9.2 ms, both P<0.0001). Using IHD anatomical models, mean QRSds were approximately 8 ms longer compared to healthy models, with slow conduction adding an additional 11 ms prolongation in the mean QRSd (Figure 1 C, D). Consistent sex-specific differences were noted.

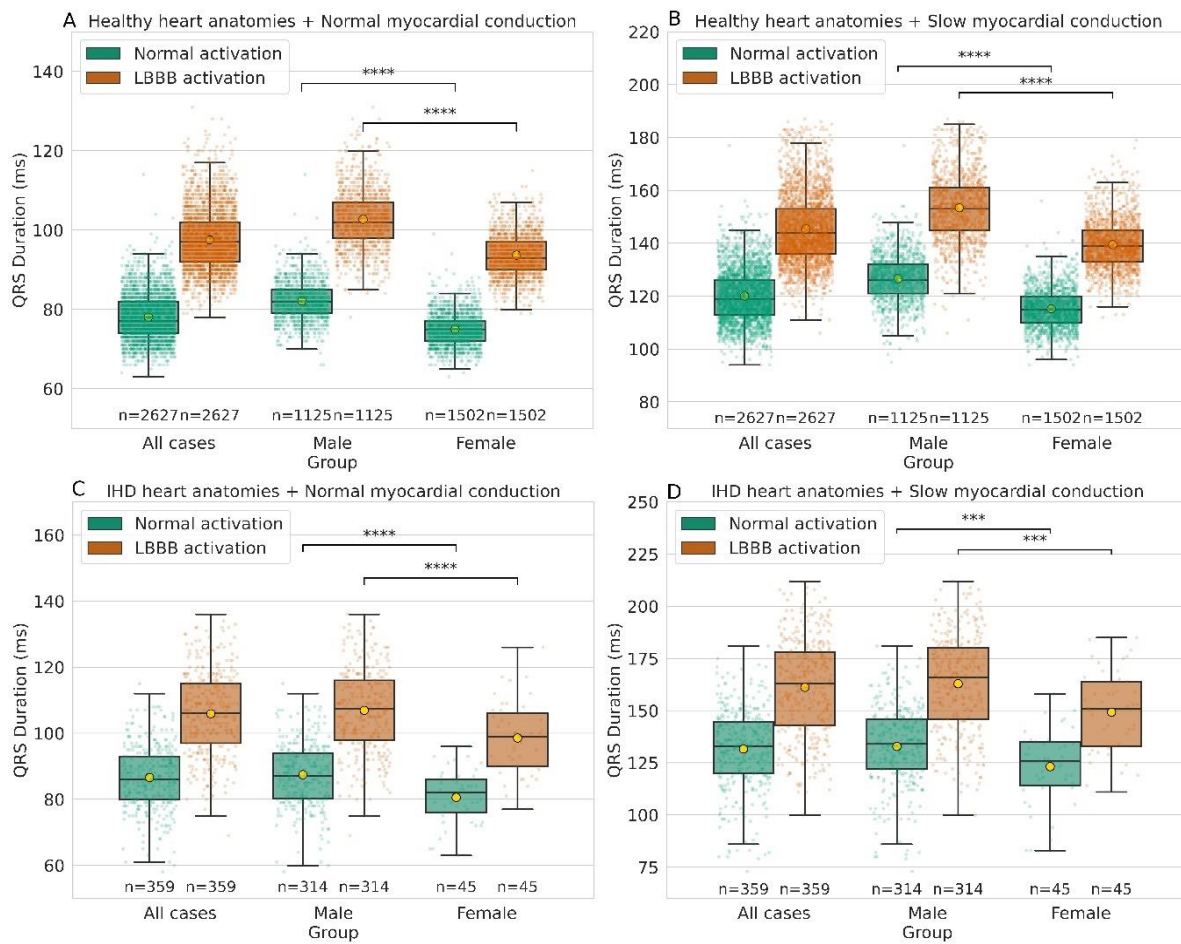

*Figure 1: Comparison of QRS durations simulated in four scenarios of both sexes in both healthy and IHD cohorts. Four scenarios include healthy or LBBB activation combined with normal or slow conduction. *P<0.05; **P<0.01; ***P<0.001; ****P<0.0001.*

## Which simulated pathological scenarios will meet the current criteria of CRT

To test how our models resemble the patients' real conditions, we selected patients from the IHD cohort who met the Class 1A indication for CRT ($LVEF \leq 35\%$ and $QRS\ duration \geq 150\ ms$) (n=17), and assessed whether their simulated QRSds in various scenarios still exceeded the 150 ms threshold, reflecting their realistic pathology. As shown in Figure 2, none of their simulated QRSds surpassed 150 ms under normal conduction scenarios, regardless of LBBB or normal activation ($113.2 \pm 13.3\ ms\ and\ 90.6 \pm 11.5\ ms$). Under slow conduction scenarios, QRSds increased with 35% and 88% of cases surpassing the 150 ms threshold (normal activation: $134.9 \pm 24.5\ ms$ and LBBB activation: $172.2 \pm 20.9\ ms$). Based on these results, we can infer that most patients who currently meet CRT criteria are likely have a combination of slowed CV and LBBB while a small amount of CRT patients may only have slow conduction.

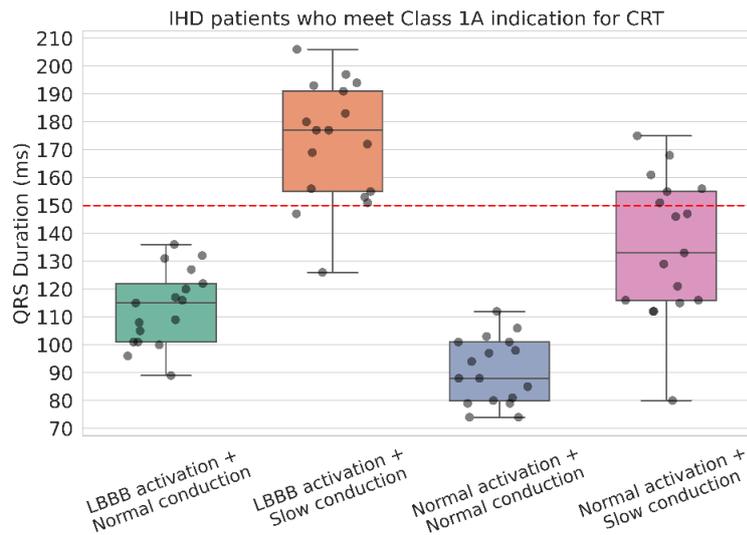

*Figure 2: Simulated QRS durations in four scenarios (LBBB or normal activation and normal or slow conduction) using the IHD heart anatomies built from patients who met the Class 1A indication of CRT (LBBB with $QRSd \geq 150\ ms,\ LVEF \leq 35\%$)(n=17)*

## LBBB and non-LBBB Patient stratification by conventional and adjusted QRSd criteria across sexes

Prior studies suggesting indexed QRSd for improving the identification of female CRT responders often analysed cohorts that already met conventional CRT criteria. However, if these indexed criteria were applied broadly to all heart failure patients, it might identify individuals who do not meet current inclusion criteria for CRT. To evaluate the true predictive accuracy in LBBB patients and 'false positive' rates in non-LBBB patients of these QRSd criteria, we apply them to a synthetic model cohort to assess their effectiveness in a general population.

Firstly, the conventional QRSd thresholds (150ms [8,9], 130 ms [8] and 120 ms [9]) were evaluated for their effectiveness in stratifying male and female patients with and without LBBB. For LBBB patients with slow conduction (Figure 3A, B), the 150 ms threshold identified 71% and 58% of males with IHD and healthy anatomies, respectively, significantly exceeding the 53% and 11% rates for females (both P<0.03). Lowering the threshold improved the true positive rate for LBBB females and achieved a non-significant sex difference when using a threshold of 120 ms (IHD: 91% and healthy: 99%), to levels comparable to males (IHD: 96% and healthy: 99%, P=0.77 and P=0.12). However, the lower thresholds also increased false positive rates among non-LBBB patients with slow conduction (Figure 3C, D), particularly in males (IHD: 76%, healthy: 77%), compared to females (IHD: 57%, healthy: 21%, both P<0.02). No significant sex differences were observed in normal conduction scenarios for either LBBB or non-LBBB patients.

Next, the QRSd normalized by heart structural metrics such as LVEDV and LV mass were assessed, using cutoff values: 0.52 and 0.65 ms/mL for LVEDV, and 0.8 ms/g for LV mass, as proposed in previous clinical trials[4,10,11]. We aimed to test if these indexed QRSd could better identify patients with LBBB, regardless of sex or slow CV. As shown in Figure 3A, B, indexed QRSd was better at identifying LBBB patients in cases with normal conduction scenarios, where (range: [21.7–99.9%]) of cases met the criteria, compared to conventional QRSd criteria ([0–12.7%]). Among them, LBBB females consistently exhibited better true predictive accuracy than males in normal conduction scenarios ([42.2-99.9%] vs [21.7-95.1%], all P<0.05). While the indexed criteria also exhibit greater false positive rate in woman (Figure 3C, D), particularly in the normal conduction scenarios ([22.2-

92.6%] vs [7.0-57.2%], all P<0.05).  In contrast, in slow conduction scenarios, indexed QRSd criteria lead to both much higher true positive rates for LBBB ([73.2-100%]) and false positive rate for non-LBBB patient selection ([49.7-100%]) with significant sex differences in most cases. Notable exceptions included the 0.52 ms/mL cutoff, which showed no significant sex differences for LBBB patients with healthy or IHD anatomies (healthy and IHD females: 100% and 93.3% vs males: 99.9% and 87.9%, P=0.43 and P=0.41) and the 0.8 ms/g cutoff in LBBB patients with IHD anatomies (female: 93.3% vs 83.8%, P=0.1).

Lastly, we assessed the diagnostic efficacy of indexed QRSd criteria by physical measures such as height using cutoff values of 0.8 and 0.9 ms/cm as proposed previously[12]. As shown in Figure 3, both cutoff values demonstrated comparable patient selection rates between sexes across most scenarios, with <10% difference observed in healthy anatomies and slow conduction scenarios using the 0.9 ms/m (P<0.05). The highest selection rates for LBBB patients occurred in slow conduction scenarios with healthy anatomies (83.7 and 83.9% for females) and IHD anatomies (83.1 and 82.2% for females) using the 0.8 ms/m cutoff. In contrast, the corresponding false positive rates of non-LBBB patients with healthy and IHD anatomies (Healthy: 37.8 and 37.6% for females; IHD: 5.6 and 2.2% for females) are lower, compared to QRSd threshold of 150 ms and all indexed QRS duration thresholds by LVEDV and LV mass.

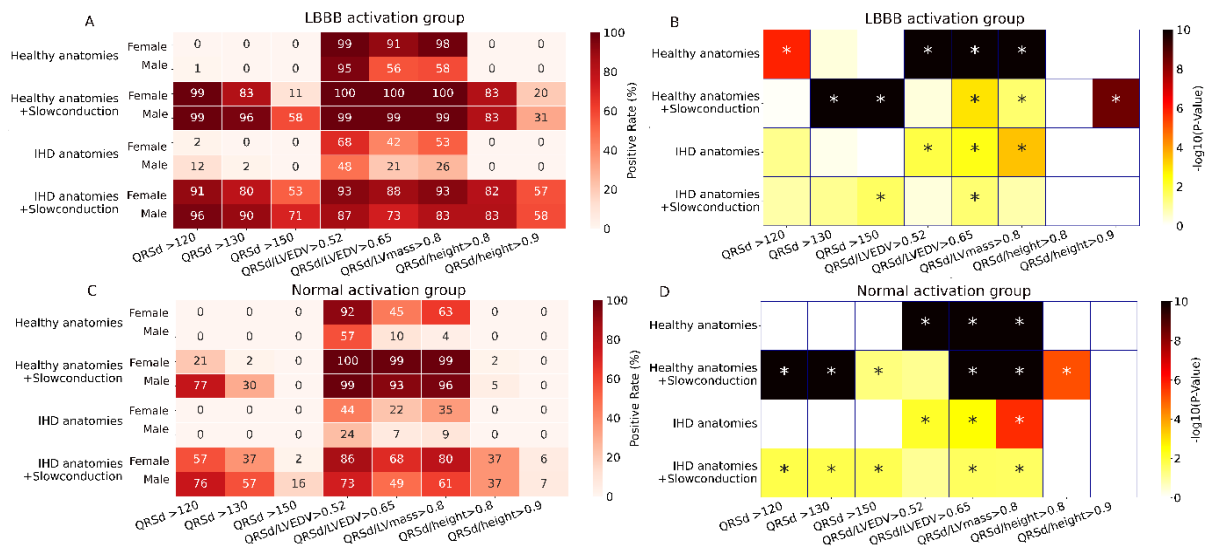

*Figure 3: Heatmaps for classification success (positive) rates for LBBB (A) and non-LBBB Patients (C) using conventional and adjusted QRSd criteria by LVEDV, LV mass and height in different sexes for four simulated scenarios including using healthy and IHD heart anatomies with normal and slow conduction. Each cell represents the success (positive) rate in %. Panel B and D represent the corresponding P values of the comparison of the success rate between male and female, for LBBB and non-LBBB activation groups. * P<0.05.*

## Classification performance of QRSd criteria for LBBB and non-LBBB patients

Prior studies have shown that patients with LBBB generally derive greater benefit from CRT than those without[6]. To distinguish LBBB patients, different criteria based on QRS morphology in the electrocardiograms have been proposed, which lead to significant variability in patient stratification[7]. Here we aim to determine whether LBBB patients can be effectively stratified using only QRSd-based criteria, rather than relying on the complex and inconsistent ECG morphology-based criteria.

To investigate the classification performance of both conventional and adjusted QRSd criteria in predicting LBBB patients under various pathological scenarios, we have plotted the ROC curves for each individual QRSd criterion mentioned above and also for both female and male groups to investigate the sex-specific difference (Figure 4). In general, the AUC for QRSd decreased as the levels of pathological changes increased, for example for the models trained on all cases: healthy anatomies+ normal conduction: 0.985> healthy anatomies + slow conduction: 0.95 > IHD anatomies + normal conduction: 0.886 > IHD anatomies + slow conduction: 0.832, so did for all the other indexed QRSd criteria, no matter of specific sex groups. Overall, the conventional QRSd model consistently outperformed those indexed by LVEDV and LV mass in all scenarios (all P<$1 \times 10^{-5}$). Interestingly, the model indexed by height demonstrated better performance overall, outperforming indexed LVEDV and LV mass models across all pathological scenarios (all P<$1 \times 10^{-5}$) while only exceeding QRSd model significantly in two scenarios with healthy anatomies (AUC: normal conduction: 0.994 vs 0.985, P=$1 \times 10^{-5}$ and slow conduction: 0.972 vs 0.95, P<$1 \times 10^{-5}$). For the models trained on specific sex groups, the AUCs of conventional QRSd and QRSd indexed by heights are similar which both show better predictive power than the QRSd indexed by LVEDV and LV mass.

The optimal thresholds of QRSd increased while the levels of pathological conditions increased, ranging from 87 ms for healthy anatomies + normal conduction to 152 ms for IHD anatomies + slow conduction. Under the IHD anatomies and slow conduction scenario, the optimal threshold of 152 ms for QRSd achieves sensitivity of 0.68 and specificity of 0.87, closely matching the height indexed threshold of 0.86 ms/m with sensitivity of 0.71 and specificity of 0.84. While in the other three scenarios with less remodeling, the height-indexed threshold achieved the best sensitivity and specificity balance, slightly better than QRSd and surpassing QRSd indexed by LVEDV and LV mass.

The male optimal thresholds of QRSd are always larger than the females with differences ranging from 7ms for healthy anatomies+ normal conduction to 13 ms for healthy anatomies+ normal conduction. In contrast, the sex-specific differences of optimal thresholds of QRSd indexed by heights are very small (varying by 0.01) under all scenarios, comparing to much larger variations as in QRSd indexed by LVEDV and LV mass (ranges: [0.02, 0.29]).

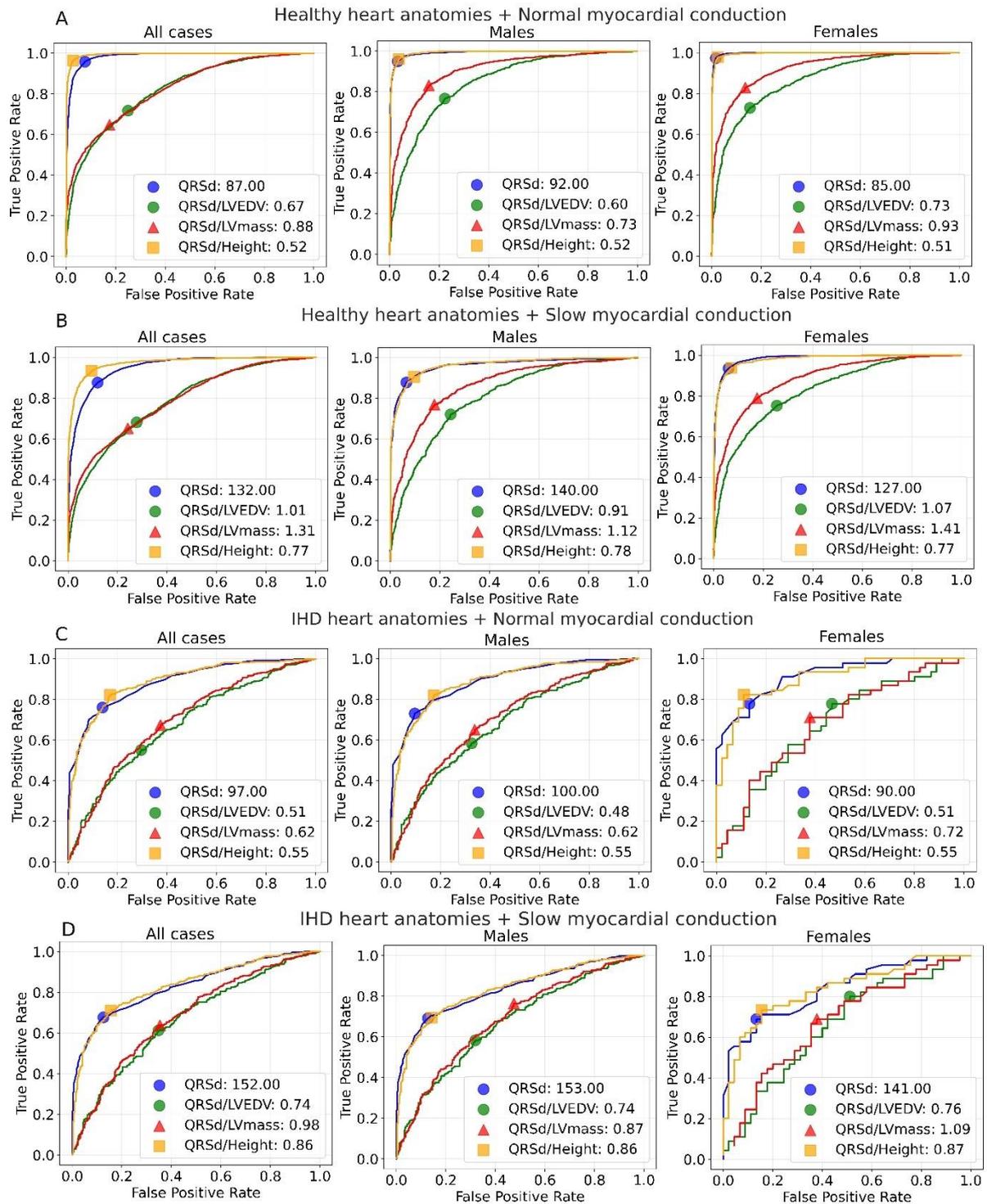

Figure 4: Classification performance assessment on conventional QRS duration and adjusted QRS duration by LVEDV, LV mass and height for classifying LBBB patientss from non-LBBB patients for different sex groups. Four pathological scenarios are included in this analysis which are healthy and IHD anatomies with normal (A,C) and slow conduction (B,D). The markers in different shapes represent the optimal threshold for the coresponding predictors.

# Discussion

This study, built on large virtual cohorts, highlights the interplay between structural remodeling, LBBB and conduction abnormalities in influencing QRSd, therefore leads to distinct diagnostic efficacy of conventional and indexed QRSd criteria in stratifying CRT patients with LBBB. We have observed the well-known sex differences in QRSd across different pathological scenarios. Through the virtual LBBB patient selection experiments, we observed significant sex differences in diagnostic efficacy of conventional QRSd thresholds (120, 130 and 150 ms), potentially explaining better CRT outcomes in females, particularly those with shorter QRSds. Indexing by LVEDV and LV mass mitigated some sex differences but significantly increased the 'false' positive rates for non-LBBB patients, while indexing by height effectively resolved sex disparities while maintaining low non-LBBB selection rates, making it the most reliable predictor for CRT patient selection. Compared to other structural metrics, indexing to height is a simple and quick method using data that is routinely measured in current practice, therefore could be easily tested and validated at a large scale, offering huge potential for widespread clinical applications.

Randomized controlled trials (RCTs) have been instrumental in establishing CRT guidelines, but they have notable challenges in addressing equity and diversity. For example, in the current RCTs, recruited patients are predominantly northern European origin with a female minority (~27%) [3,12]. Furthermore, discrepancies in the definitions of LBBB morphology between American and European guidelines, as well as across different guideline versions—such as ESC-2021 and ESC-2013—have been demonstrated to result in significant variations in patient stratification[7] and inconsistencies in specificity, particularly in the context of structural remodeling[13]. In contrast, our modelling approach offers an unbiased methodology for evaluating CRT patient selection, distinct from the conventional analysis on RCTs, which only included patients meeting the CRT clinical guideline inclusion criteria. This approach enables precise 'control' over pathological substrates for each virtual patient, replicating the 'true LBBB' condition while integrating known structural and other functional remodelling such as myocardial conduction delay. By creating a comprehensive population-based library of virtual patients, our method overcomes the biases inherent in retrospective RCT-based analyses and provides a robust in-silico platform for assessing and refining CRT patient selection criteria.

In order to find the most representative simulated pathological scenarios performed here, we performed a subset analysis on the group of 17 IHD patients who met the Class 1A indication for CRT (LVEF ≤ 35% and QRS duration ≥ 150 ms) based on their clinical records. We found that a majority of cases (15 of 17) simulated under the LBBB and slow conduction scenario exceeded the 150 ms threshold, while 6 of 17 cases simulated under LBBB and normal conduction exceeded the 150ms threshold. These results inferred that most of the patients who met the Class 1A criteria are likely to have a combination of LBBB and slowed CV, suggesting our models effectively replicate different degrees of cardiac remodelling, aligning with clinical observations. Furthermore, our ROC analysis on stratifying LBBB and non-LBBB patients identified an optimal QRSd threshold of 152 ms in IHD patients with slow conduction, closely matching current guidelines and reinforcing the credibility of our approach.

Consistent with previous observed clinical results[14,15], our simulation results showed significant sex differences in the baseline QRSds in both cohorts with/without structural remodeling. Added functional remodeling, such as slow myocardial conduction and LBBB dysfunction, only amplified these differences across sexes. The significant influence of heart size is also evident in our findings, as the IHD patients who have substantially larger hearts (mean LVEDV: 215.9 (IHD) vs 134.9 mL) exhibited longer baseline QRSds, compared to the healthy participants (mean: 86.6 (IHD) vs 78.0 ms). These results align with previous studies[16,17] highlighting the role of sex-specific heart sizes and functional remodeling in influencing the QRSds.

Our modelling results revealed a significant sex difference in the diagnostic efficacy of the conventional QRSd thresholds of 120, 130 and 150 ms. The 150 ms QRSd under-selected more LBBB females for CRT than males, no matter with or without structural remodeling. It suggests that females selected at this threshold might represent a more severely affected subgroup, potentially explaining their superior CRT outcomes, as shown in previous studies[12,16]. In contrast, reducing to 120 ms substantially improves LBBB patient selection for both sexes and eliminates significant sex disparities. However, this adjustment also raised non-LBBB selection rates, particularly in males. This increased patient selection in LBBB females and non-LBBB males may jointly contribute to the clinically observed better outcomes in females with narrower QRSd thresholds[12,16] and highlighted a trade-off between sensitivity and specificity in CRT candidate selection.

Indexing QRSd by cardiac structural indices such as LVEDV and LV mass partially addressed sex disparities in QRSd. While these adjustments improved sensitivity by identifying more LBBB patients, the concurrent increase in non-LBBB selection rates led to reduced specificity, limiting their overall diagnostic performance. Additionally, significant sex differences persisted in most pathological scenarios, suggesting that these structural metrics may not fully account for intrinsic sex-based differences in heart size and remodelling patterns. This was reaffirmed in the ROC analysis that the AUCs for indexing by LVEDV and LV mass are lower than QRSd in any pathological scenarios.

Height-indexed QRSd criteria emerged as the most effective approach, resolving sex disparities in LBBB patient selection across most pathological scenarios as shown in selected threshold assessments (Figure 5), while maintaining relatively low non-LBBB selection rates. Moreover, our ROC analysis further confirmed the superior classification performance of LBBB patients using height-indexed criteria than the other QRSd criteria. In the separate sex-specific analysis, we further observed very similar optimal thresholds of height-indexed QRSd for both sex groups, reaffirming its power to resolve the sex disparities. These criteria leverage height, as a stable biomarker closely associated with baseline heart size, avoiding the variability introduced by pathological remodelling and aging, which potentially address the intrinsic heart size differences between sexes.

This study provides a novel population-based modelling approach, advanced from previous similar modelling work with a small patient cohort (<10)[13], to evaluate CRT patient selection criteria, free from biases inherent in clinical trial recruitment. Unlike traditional RCTs, our approach simulates controlled pathological scenarios, enabling the assessment of QRS duration criteria across diverse patient populations. Importantly, this method captures the full spectrum of LBBB, structural remodelling, and conduction abnormalities, offering a comprehensive framework for refining CRT guidelines.

Our findings reveal substantial sex differences in QRS duration and the diagnostic efficacy of conventional thresholds. While indexing QRSd by cardiac structural metrics such as LVEDV and LV mass improved sensitivity, it also reduced specificity and failed to eliminate sex disparities. In contrast, height-indexed criteria effectively resolved these differences while maintaining diagnostic accuracy, emerging as the most reliable predictor for CRT candidate selection. These results uncover a key mechanistic driver of sex differences in CRT outcomes and underscore the importance of incorporating sex-specific considerations into updated CRT guidelines to ensure equitable and effective treatment.

# Methods

## Study population and mesh generation

This study consists of a cohort of healthy participants and a clinical cohort of IHD patients. The healthy cohort is collected from the UK biobank (UKBB) (https://www.ukbiobank.ac.uk/), a large-scale prospective study of middle/old age individuals in the UK initiated in 2006. From an initial cohort of 3461 participants[5] , 2627 participants with no cardiovascular disease (CVD) were included in this study. Participants with any diagnosis of 'Diseases of the Circulatory system' (I00-I99) according to the International Classification of Diseases- 10th Edition (ICD-10) (Data-Field 41270) were excluded. The clinical cohort of patients with IHD (n=359), was recruited at the Royal Brompton & Harefield NHS hospital as previously reported[18].

Patient-specific biventricular meshes were generated from cardiac magnetic resonance (CMR) images for both cohorts as described previously [5]. In brief, the CMR images taken at the end-diastolic phase were automatically segmented to derive contours and landmarks of both ventricles. They were then processed using an established atlas-based pipeline to generate surface meshes, which were subsequently converted to volumetric meshes.

To enable consistency in simulation setups, we implemented a morphological coordinate system, also known as universal ventricular coordinates (UVCs), for describing positions within ventricles based on the apical-basal ($Z$), transmural ($\rho$) (from endocardium to epicardium), rotational ($\varphi$) (anterior, anteroseptal, inferior, inferolateral, anterolateral) and chamber-wise (left ventricle and right ventricle) coordinates, and applied to all models. Biventricular myocardial fibre structure was incorporated using a rule-based method with transmural angle ranging from 60° to -60° longitudinally and −65° to 25° transversely, transitioning from endocardium to epicardium. Details regarding population sample selection and mesh quality control have been described previously [5].

## Electrophysiological (EP) simulations

The electrophysiological simulations were performed using the cardiac arrhythmia research package (CARP)[19]. A Reaction-Eikonal model without diffusion was used to compute the ventricular activation times and transmembrane potential dynamics[20]. Similar to previous studies[5,21], the ventricular depolarization during sinus rhythm was initiated using a fascicular-based model to replicate the His-Purkinje activation. Specifically, as shown in Figure 5, there are five early activation sites located as follows: three in the left ventricle (LV)-anterior (af), septal (sf) and posterior(pf) walls, and two in the right ventricle (RV)-septal wall (sf) and the lateral moderator band (mod). Early activation sites are represented as 20 $\mu m$ radius disks (~25 cells) defined by the universal apical-basal and rotational coordinates. A sub-endocardial (SE) layer with faster conduction was implemented as a thin layer near the endocardium, defined by transmural coordinates $\rho_{SE}$ and apical-basal coordinates $Z_{min,SE}, Z_{max,SE}$. All parameters mentioned above are shown in Table 2, and their values were set as the median values derived from previous measurements reported in the literature.

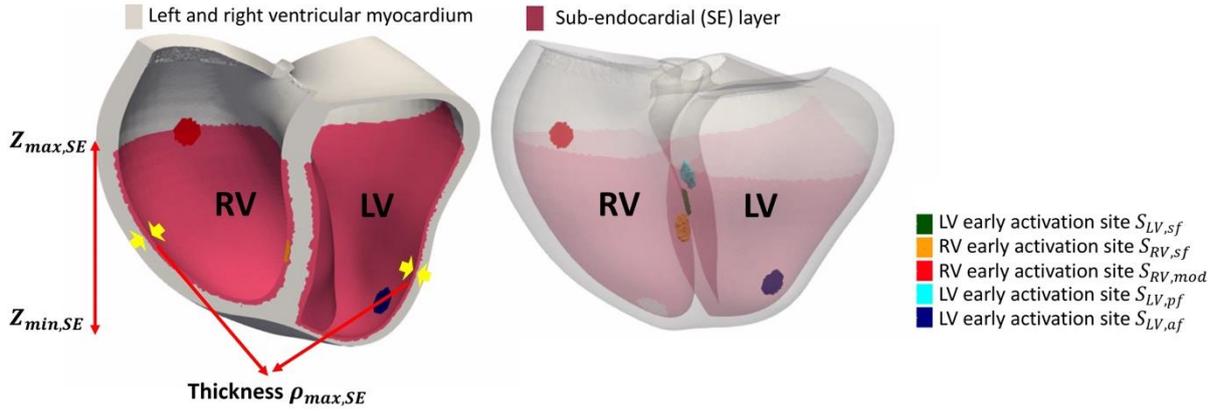

*Figure 5: The fascicular-based model replicates the realistic ventricular activation modulating by the His-Purkinje system.*

| Parameters | Value | Description | References |
|---|---|---|---|
| $Z_{LV,sf}$ | 0.5 | Apical-basal coordinate for LV septal fascicle | 22–25 |
| $Z_{LV,pf}$ | 0.5 | Apical-basal coordinate for LV posterior fascicle | 22–24 |
| $Z_{LV,af}$ | 0.7 | Apical-basal coordinate for LV anterior fascicle | 22–24 |
| $Z_{RV,sf}$ | 0.6 | Apical-basal coordinate for RV septal fascicle | 22–24 |
| $Z_{RV,mod}$ | 0.4 | Apical-basal coordinate for RV moderate band fascicle | 22–24 |
| $\varphi_{LV,sf}$ | 0 | Rotational coordinate for LV septal fascicle | 22–25 |
| $\varphi_{LV,pf}$ | -1 | Rotational coordinate for LV posterior fascicle | 22–25 |
| $\varphi_{LV,af}$ | 2.37 | Rotational coordinate for LV anterior fascicle | 22–25 |
| $\varphi_{RV,sf}$ | 1 | Rotational coordinate for RV septal fascicle | 22–25 |
| $\varphi_{RV,mod}$ | 0 | Rotational coordinate for RV moderate band fascicle | 22–25 |
| $t_{LV,sf}$ | 15 | Firing time for LV posterior fascicle | 22 |
| $t_{LV,pf}$ | 15 | Firing time for LV anterior fascicle | 22 |
| $t_{LV,af}$ | 15 | Firing time for RV septal fascicle | 22 |
| $t_{RV,sf}$ | 15 | Firing time for RV moderate band fascicle | 22 |
| $t_{RV,mod}$ | 15 | Firing time for LV posterior fascicle | 22 |
| $\rho_{SE}$ | 0.155 | Transmural coordinate for SE layer | 21,26 |
| $Z_{min,SE}$ | 0.08 | The minimum apical-basal coordinate for SE layer | 25 |
| $Z_{max,SE}$ | 0.75 | The maximum apical-basal coordinate for SE layer | 25,27,28 |
| $R_{myo,SE}$ | 5 | The ratio of CV in SE layer to CV in myocardium | 21,26,29,30 |

*Table 2: The electrophysiological parameters used for defining the fascicular-based model.*

Myocardium was simulated as a transversely isotropic medium with a conduction velocity of 0.65 m/s along fiber directions and an anisotropy ratio of 0.42 for the transverse fiber direction[31]. Normal sinus rhythm activation was simulated by stimulating all five early activation sites, while the LBBB activation was simulated by stimulating only two RV sites, as shown in Figure 6. Given that the torso information is unavailable, the heart models were registered to a heart enclosed in an existing torso model[32] using the UVCs. Within this reference torso, the locations of electrodes used in measuring 12-lead ECGs were identified, the corresponding extracellular potentials were simulated, and the ECGs were computed. The QRS duration was computed by finding the time points at which the spatial velocity exceeds 0.15 of the maximum spatial velocity in the reconstructed corresponding vectorcardiogram (VCG) from 12-lead ECGs, which fuse the information in ECG traces[32].

Slowed cardiac conduction, representing pathological processes such as ionic remodeling, gap junction uncoupling, and fibrosis following ischemia [33], was simulated by reducing the model conduction velocity to 0.4 m/s, based on previous experimental and modelling studies[34,35]. For each patient-specific anatomical model, four scenarios were simulated: (1) normal activation with normal conduction, (2) normal activation with slowed conduction, (3) LBBB activation with normal conduction, and (4) LBBB activation with slowed conduction.

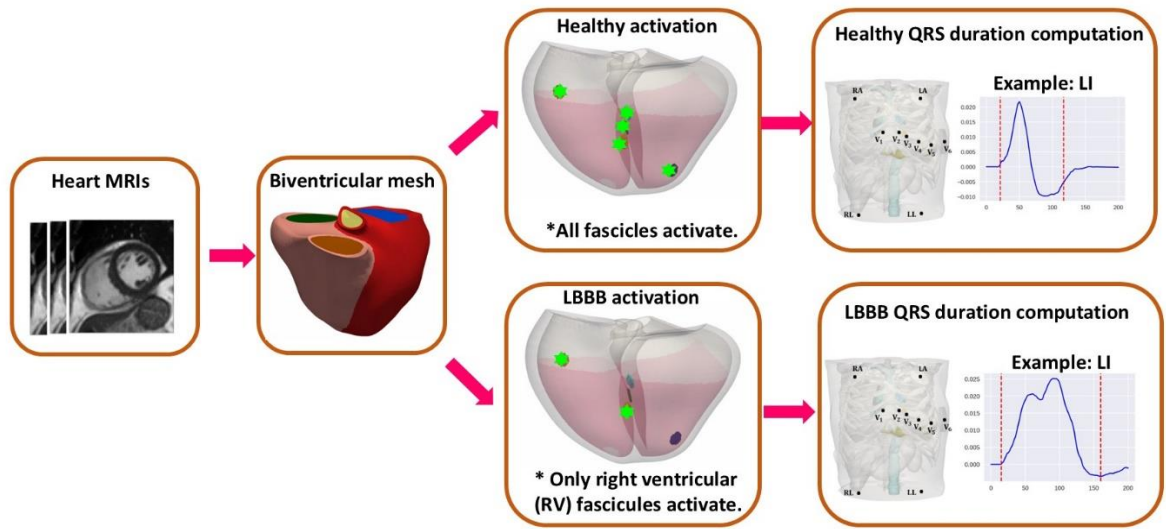

*Figure 6: The workflow of this study. Each patient-specific biventricular mesh was activated by all fascicles representing healthy condition and only right ventricular fascicles representing left bundle branch block. The resultant QRS durations were computed from the reconstructed 12 lead ECGs.*

## Statistical analysis

Continuous variables are presented as the mean ± standard deviation and compared using the Mann-Whitney U-test, a non-parametric statistical test, due to the non-normal data distribution. Categorical variables are presented as percentages and compared using the Chi-square test or Fisher exact test, as appropriate. The receiver operating characteristic (ROC) curves were generated for all conventional and indexed QRS duration criteria across simulated scenarios. Optimal thresholds were determined using Youden's J statistics. The Delong test was used to compare the area under the receiver operating characteristic (AUC) of any pairs of correlated predictors.

# Data availability

The imaging data and non-imaging participant phenotypes and clinical outcomes are available from the UK Biobank via a standard application procedure at http://www.ukbiobank.ac.uk/register-apply.

# Code availability

The anatomical model generation pipeline is available at https://www.github.com/cdttk/biv-volumetric-meshing, which includes the use of open-source software: openCARP, available at https://opencarp.org/. The pipeline of electrophysiological simulations presented in this study will be made publicly available once accepted. These electrophysiological simulations were performed using the Reaction-Eikonal model, which is part of a proprietary version of the Cardiac Arrhythmia Research Package (CARP)[36] and requires a license. However, this can also be performed using a monodomain model in openCARP, which is fully open source, albeit computationally more expensive.

# Acknowledgments

This research has been conducted using the UK Biobank Resource under application number 88878. It is supported by the Wellcome Trust and the Engineering and Physical Sciences Research Council (EPSRC) Centre for Medical Engineering (WT203148/Z/16/Z). S.N. is also supported by National Institutes of Health R01-HL152256, European Research Council (ERC) PREDICT-HF 453 (864055), the British Heart Foundation (BHF) (RG/20/4/34803), the EPSRC (EP/X012603/1 and EP/P01268X/1), and the Technology Missions Fund under the EPSRC grant EP/X03870X/1 and the Alan Turing Institute. M.B. acknowledges BHF project grants PG/22/11159 and PG/22/10871. E.V. is supported by French National Research Agency ANR-10-IAHU-04. G.P. is supported by the Austrian Science Fund (FWF) grant 10.55776/I6540. B.P.H. is supported by the BHF (FS/ICRF/24/26128), the Rosetrees Trust and the National Institute for Health and Care Research (NIHR) Imperial Biomedical Research Centre (BRC).

# Author contributions

S.Q., C.R., A.Y., P.L., M.B. and S.N. conceived and designed the study. S.Q., D.U., and E.F. constructed anatomical models for the UK biobank cohort. R.E.J., H.Z., S.P., B.P.H. and D.H. contributed to the data collection and quality control check of the clinical cohort of patients with ischemic heart disease, and S.Q., H.Z., A.Y. and M.B. contributed to the model construction of those patients. G.P. and E.V. contributed to the development and maintenance of the CARPEntry software used for electrophysiological simulations. S.Q. implemented the pipeline for simulating LBBB/non-LBBB scenarios, applied to both cohorts, analyzed the results and wrote the manuscript. All authors contributed to the manuscript revision and approved the final submitted version.

# Competing Interests

The authors declare no competing interests.